\documentclass[twocolumn,showpacs,preprintnumbers,prl,aps,amssymb,superscriptaddress]{revtex4-1}
\usepackage{graphicx}

\begin{document}

\newcommand{\Co}{CeCoIn$_5$}
\newcommand{\La}{Ce$_{1-x}$La$_x$CoIn$_5$}
\newcommand{\Nd}{Ce$_{1-x}$Nd$_x$CoIn$_5$}
\newcommand{\Yb}{Ce$_{1-x}$Yb$_x$CoIn$_5$}

\newcommand{\ie}{{\it i.e.}}
\newcommand{\eg}{{\it e.g.}}
\newcommand{\etal}{{\it et al.}}

%%%%%%%%%%%%%%%%%%%%%%%%%%%% TITLE

\title{Evolution of the superconducting energy gap structure concomitant with Fermi surface reconstruction in the heavy-fermion superconductor CeCoIn$_5$ }

%%%%%%%%%%%%%%%%%%%%%%%%%%%% AUTHORS

\author{Hyunsoo Kim}
\affiliation{Ames Laboratory and Department of Physics \& Astronomy, Iowa State University, IA 50011, USA}

\author{M.~A.~Tanatar}
\affiliation{Ames Laboratory and Department of Physics \& Astronomy, Iowa State University, IA 50011, USA}

\author{R.~Flint}
\affiliation{Ames Laboratory and Department of Physics \& Astronomy, Iowa State University, IA 50011, USA}

\author{C.~Petrovic}
\affiliation{Department of Physics, Brookhaven National Laboratory, Upton, New York 11973, USA}

\author{Rongwei Hu}
\altaffiliation[Present address: ] {Rutgers Center for Emergent Materials and Department of Physics and Astronomy, Rutgers University, Piscataway, New Jersey 08854, USA.}
\affiliation{Department of Physics, Brookhaven National Laboratory, Upton, New York 11973, USA}

\author{ B.~D.~White}
\affiliation{Department of Physics, University of California, San Diego, La Jolla, California 92093, USA}

\author{I.~K.~Lum}
\affiliation{Department of Physics, University of California, San Diego, La Jolla, California 92093, USA}

\author{ M.~B.~Maple }
\affiliation{Department of Physics, University of California, San Diego, La Jolla, California 92093, USA}

\author{R.~Prozorov}
\email[Corresponding author: ]{prozorov@ameslab.gov}
\affiliation{Ames Laboratory and Department of Physics \& Astronomy, Iowa State University, IA 50011, USA}

\date{14 April 2014}

%%%%%%%%%%%%%%%%%%%%%%%%%%%% ABSTRACT

\begin{abstract}

The London penetration depth, $\lambda (T)$ was measured in single crystals of Ce$_{1-x}R_x$CoIn$_5$, $R$=La, Nd and Yb down to 50~mK ($T_c/T \sim$50) using a tunnel-diode resonator.  In the cleanest samples $\Delta \lambda (T)$ is best described by the power law, $\Delta \lambda (T) \propto T^{n}$, with $n \sim 1$, consistent with line nodes. Substitutions of Ce with La, Nd and Yb lead to similar monotonic suppressions of $T_c$, however the effects on $\Delta \lambda(T)$ differ. While La and Nd doping results in an increase of the exponent to $n \sim 2$, as expected for a dirty nodal superconductor, Yb doping leads to $n > 3$, inconsistent with nodes, suggesting a change from nodal to nodeless superconductivity where Fermi surface topology changes were reported, implying that the nodal structure and Fermi surface topology are closely linked.

\end{abstract}

\maketitle

%%%%%%%%%%%%%%%%%%%%%%%%%%%% INTRODUCTION

Magnetically mediated pairing is widely believed to be responsible for unconventional superconductivity found in materials ranging from the high-T$_c$ cuprates to the iron-based superconductors\cite{Norman} to heavy fermion compounds\cite{Mathur, Monthoux,Sarrao}.  For a long time, this unconventional superconductivity was thought to always be nodal, with a d-wave superconducting energy gap symmetry.  While unconventional pairing does require a sign changing gap, nodal lines are not actually required, and many iron-based superconductors have an s$_\pm$ gap structure, where any nodes are merely accidental.  Recently there have been some suggestions of fully gapped\cite{Kittaka} or s$_\pm$ \cite{Ronning,TDas} superconductivity in heavy fermion materials.  In this paper, we use penetration depth studies to show that nodes in the superconducting energy gap of pure CeCoIn$_5$ can be \emph{removed} by substituting Yb for Ce, revealing the first clear example of a nodeless heavy fer
 mion superconductor.

The heavy fermion superconductor CeCoIn$_5$ has one of the highest transition temperatures in its class, $T_c$=2.3~K \cite{Petrovic} and reveals quantum criticality when tuned by either pressure \cite{Sidorov} or field \cite{Paglione,Movshovich-QCP,Science}. The criticality is thought to be due to magnetic fluctuations, making it an intriguing material in which to study the relationship between magnetism, quantum criticality and the superconducting energy gap structure. % also recent paper claiming zero-field QCP
Several experimental studies have shown unconventional properties consistent with the presence of line nodes in the superconducting energy gap \cite{NQR,Movshovich-unconv,Brown,PCS}. Thermal conductivity and heat capacity measurements as a function of magnetic field direction \cite{Izawa,Sakakibara} are interpreted \cite{Vekhter} as evidence for a $d_{x^2-y^2}$ gap. This interpretation finds support in directional point contact spectroscopy \cite{Laura} and $k$-space resolved quasiparticle interference scanning tunneling microscopy (STM) measurements \cite{Allan,Yazdani}, as well as the spin resonance found at a three-dimensional ($\pi$,$\pi$,$\pi$) wavevector \cite{Stock}. 

However, none of these measurements are phase sensitive and thus can only demonstrate the presence and perhaps positions of line nodes, not their origin and symmetry.  Indeed, several observations are difficult to reconcile with the $d-$wave scenario. Most importantly, despite very low residual resistivity $\rho_0$=0.2$\mu \Omega$cm in magnetic fields of 5~T \cite{nonvanishing}, the London penetration depth of pure CeCoIn$_5$ has never shown the linear temperature dependence expected in clean $d$-wave superconductors.  Instead, if $\lambda(T)$ is parametrized by fitting to a power-law, $\Delta \lambda (T) =AT^{n}$, measurements on crystals from different sources that presumably have different amounts of scattering \cite{Ormeno,Brown,Chia,PNAS} yield a variation of the exponent $n$ between 1.5 and 2, where $n=2$ represents the dirty limit in the gapless regime for any pairing symmetry \cite{Prozorov-gapless}. Similar conclusions about the presence of a large density of unconde
 nsed quasi-particles over an extremely broad temperature and field range were made from doping-dependent thermal conductivity studies \cite{uncondensed,Gorkov,Grenoble}. The origin of this unusual superfluid response in a very clean material remains unclear, and several suggestions were put forward, including non-local electrodynamics\cite{Chia} and a temperature dependent quasi-particle mass enhancement within the superconductor due to a nearby QCP \cite{Broun2,PNAS}.  Deviations from a simple $d$-wave scenario have stimulated discussions of alternative models in which the Fermi surface topology plays an important role in the superconducting pairing in the 115s\cite{Ronning,TDas}, motivated by recent ideas put forward for iron-based superconductors \cite{MazinNature,Chubukovreview}.

To gain an insight into this unusual superfluid response of CeCoIn$_5$, in this letter we report a systematic study of the London penetration depth in single crystals of CeCoIn$_5$, with Ce substituted by both magnetic and non-magnetic rare-earth ions: La, Nd and Yb.
These three dopants affect the parent material differently: La acts as a non-magnetic impurity; excess $f$-electrons on Nd ions remain localized and induces long range magnetic order in compositions $x\geq$0.05, with $T_N <Tc$ \cite{CedoNd,Ndneutrons}; and Yb substitution provides hole doping, leading to a change in the Fermi surface topology \cite{Polyakov,Dudy}.  % need to add some comments about different types of disorder and the effect on Tc - hint already that Yb is special (well, they don't believe this is conclusive).  Also, must make that claim on the first page if we want it to get into PRL

We find that the evolution of the low - temperature variation of the London penetration depth with La and Nd substitutions is consistent with the presence of line-nodes, while its evolution with the Yb substitution suggests a nodal to nodeless transformation of the superconducting energy gap structure concomitant with the Fermi surface topology change, a challenge for the conventional d-wave picture of magnetically mediated pairing.

%%%%%%%%%%%%%%%%%%%%%%%%%%%% ExPERIMENTAL

%

Single crystals of Ce$_{1-x}$R$_x$CoIn$_5$ ($R$=La, Nd, Yb) were grown using In flux method \cite{CedoNd,PetrovicLa,Capan,MapleYb}. The values of $x$ were in the range of $0<x<0.05$ for $R$=La and Nd and $0<x<0.20$ for Yb. Samples for in-plane London penetration depth measurements were cut and polished into rectangular parallelepipeds with typical dimensions $\sim 0.6 \times 0.6 \times 0.1$~mm$^3$ ($a\times b\times c$). Details of the TDR measurements of London penetration depth in a dilution refrigerator and their analysis can be found elsewhere \cite{SrPd,Prozorov2000,Prozorov2006}.

%%%%%%%%%%%%%%%%%%%%%%%%%%%% RESULTS AND DISCUSSION

%%%%%%%%%%%%% figure 1: rf susceptibility VS. T
\begin{figure}
\centering
\includegraphics[totalheight=4in]{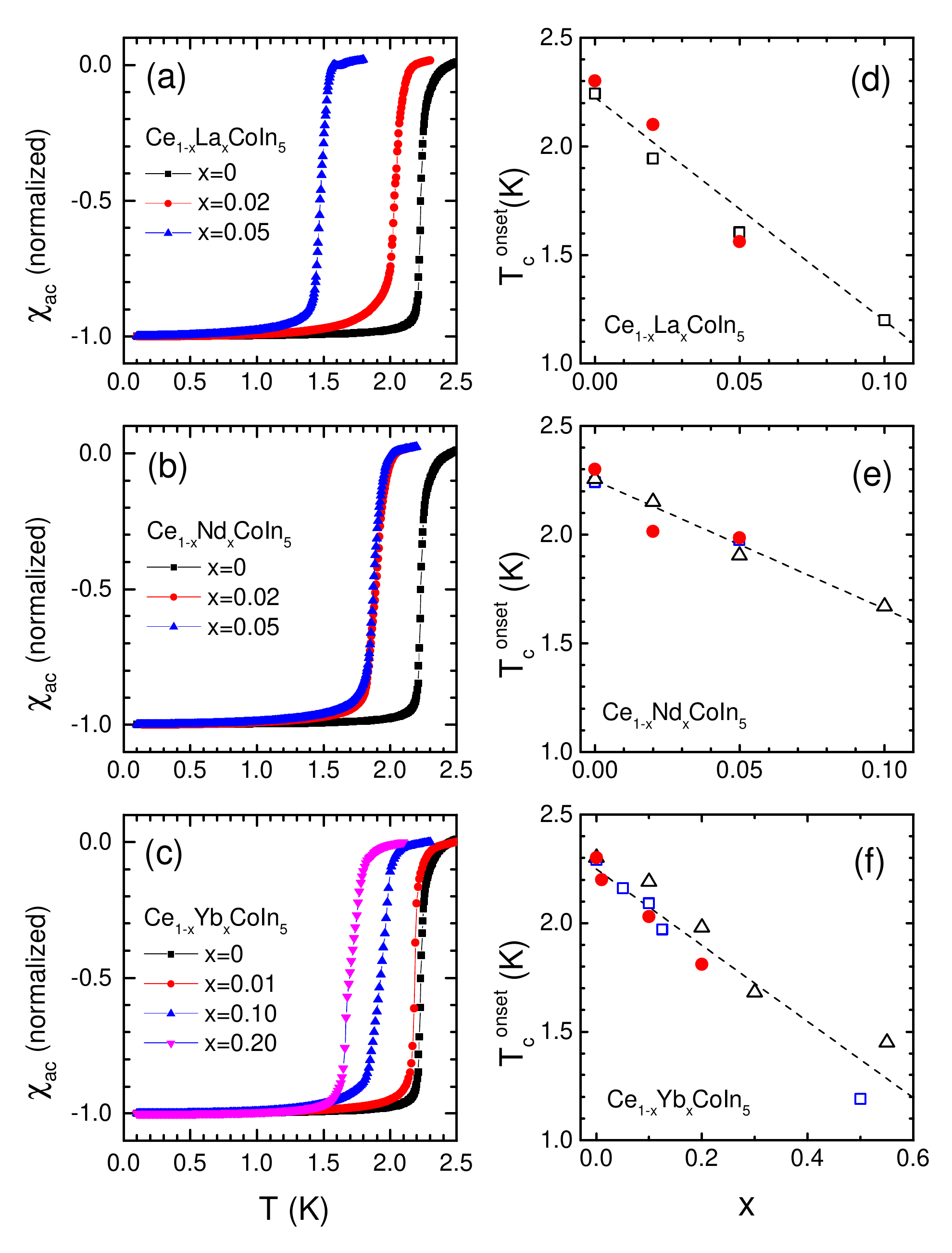}
\caption{\label{fig1_samples} 
(Color online) Left row panels (a) to (c) show the temperature-dependence of normalized rf magnetic susceptibility of Ce$_{1-x}$R$_x$CoIn$_5$ for $R$=La (top panel (a), $x$=0, 0.02 and 0.05 right to left), $R$=Nd (middle panel (b), $x$=0, 0.02 and 0.05 right to left) and $x$= Yb (bottom panel (c), $x$=0, 0.01, 0.1 and 0.2, right to left). Right row panels (d) to (f) show $T_c(x)$ as determined in our measurements (red solid dots) in comparison with the literature data for Ce$_{1-x}$La$_x$CoIn$_5$ (panel (d), data from Petrovic {\it et al.} \cite{PetrovicLa}), Ce$_{1-x}$Nd$_x$CoIn$_5$ (panel (e), data from Petrovic {\it et al.} \cite{CedoNd}) and  Ce$_{1-x}$Yb$_x$CoIn$_5$ (panel (f), blue squares are data from Capan {\it et al.} \cite{Capan} and black triangles are data from Shu {\it et al.} \cite{MapleYb}). 
}
\end{figure}

%%%Sample characterization

Figure \ref{fig1_samples} shows the temperature-dependent normalized rf magnetic susceptibility of the samples used in this study over the whole superconducting range from base temperature to $T_c$. In all cases, chemical substitution suppresses $T_c$, with $T_c(x)$ in good quantitative agreement with previous studies \cite{CedoNd,PetrovicLa,Capan,MapleYb} as shown in Fig.~\ref{fig1_samples} (d)-(f). The transitions remain sharp even in doped samples, suggesting homogeneous dopant distribution.

%%%%%%%%%%%%% figure 2: low temperature nodal vs full gap
\begin{figure}
\centering
\includegraphics[totalheight=2.7in]{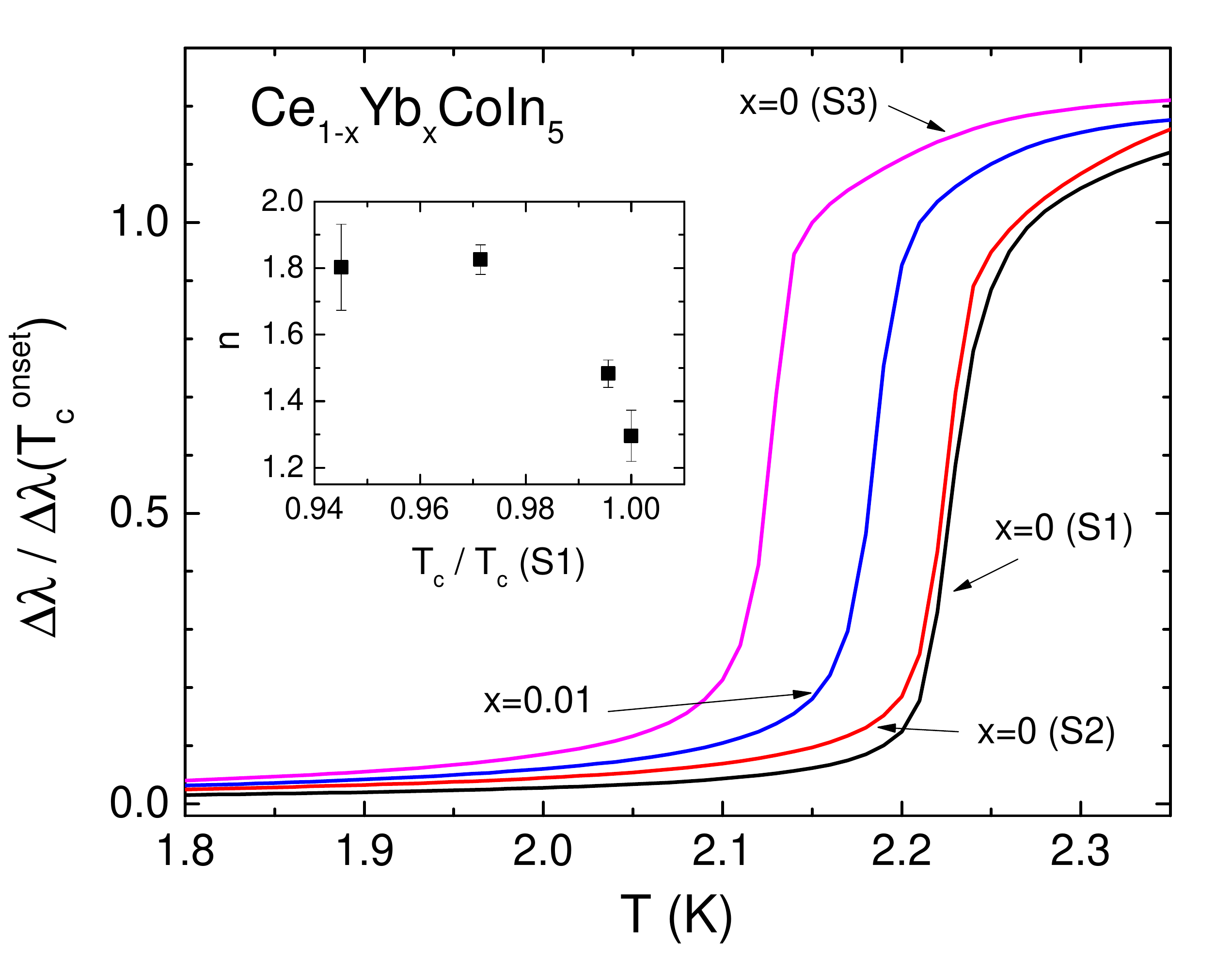}
\caption{\label{fig2_purity} 
Temperature dependence of London penetration depth in three nominally pure samples of CeCoIn$_5$ (S1, S2, and S3) and in a slightly Yb doped sample $x$=0.01. Note that S3 of the nominally pure CeCoIn$_5$ has $T_c$ lower than the Yb-doped sample. The exponent $n$, describing the temperature dependence of the London penetration depth as $\Delta \lambda (T)=AT^n$ strongly depends on $T_c$, tending to $n$=1 in the best samples.}
\end{figure}

In Fig.~\ref{fig2_purity} we show the temperature-dependent London penetration depth in a range close to $T_c$ in three nominally pure samples of CeCoIn$_5$, S1, S2 and S3. For reference we show measurements made in slightly Yb doped sample, $x$=0.01, with all measurements taken in identical conditions in the same setup and using the same thermometry. This comparison clearly shows that the $T_c$ of nominally pure samples varies by as much as 0.1~K, reflecting hidden disorder/chemical contamination. Not unexpectedly, the temperature dependence of the London penetration depth, taken in the lowest temperature limit, changes with $T_c$. Fitting data with a power-law function, $\Delta \lambda (T)=AT^{n}$, we find that $n$ sensitively depends on minute variation of sample $T_c$, as shown in the inset of Fig.~\ref{fig2_purity}. In the highest $T_c$ sample (S1), the exponent $n=$1.25 is below 1.5 and close to 1, as expected for superconductors with line nodes in the clean limit. We u
 se the data for this sample for reference in the following. For sample S3 and the Yb-doped sample ($x$=0.01) the exponent is significantly higher, tending toward $n$=2. Both $n$ close to 1 and its tendency towards $n$=2 with disorder are in line with expectation for nodal superconductors. 
For example, in the $d$-wave case confirmed by phase-sensitive experiments in cuprates \cite{KirtleyTsuei} and strongly suggested but not confirmed in CeCoIn$_5$ \cite{Izawa,Sakakibara,Vekhter,Laura,Allan,Yazdani}, the dependence is expected to be $T$-linear in the clean limit and $T^2$ in the dirty limit \cite{HirschfeldGoldenfeld}. The dependence at intermediate dopings is described by $\Delta \lambda(T) = At^2/(t^*+t)$ where $t^*$ is a crossover temperature scale determined by unitary-limit impurity scattering and $t \equiv T/T_c$.
The change of exponent is unexpectedly steep with $T_c$, suggesting that the effect of disorder is not insignificant in nominally pure CeCoIn$_5$, and might be responsible for the unusual exponents found in previous studies.

%%%%%%%%%%%%% figure 3: doping dependence of LPD
\begin{figure*}
\centering
\includegraphics[totalheight=2.9in]{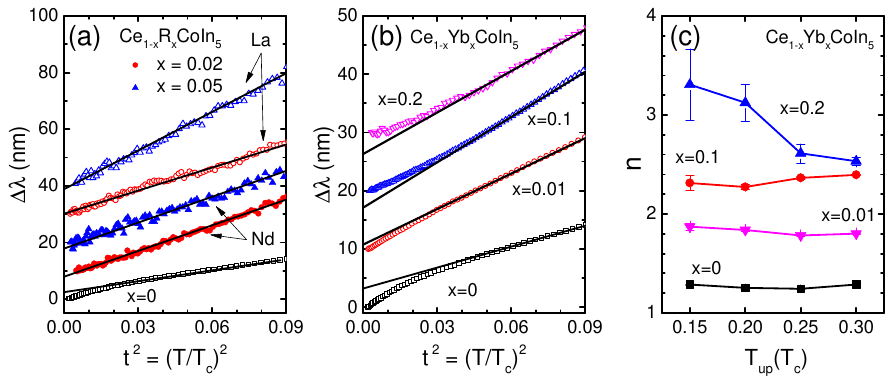}
\caption{\label{fig3_doping} 
Temperature-dependent London penetration depth of (a) La-, Nd- and (b) Yb-substituted substituted CeCoIn$_5$, plotted vs a normalized $(T/T_c)^2$ scale. The dependence in pure material S1 shows a clear downturn consistent with $n=1.2<2$.  The data for La-and Nd-doped samples closely follow a $T^2$ dependence, expected in dirty nodal superconductors  for all doping levels. In Yb-substituted samples, there is a clear crossover from sub-linear to super-linear, suggesting a rapid increase of the exponent $n$, and $n>2$ for samples with $x$=0.1 and 0.2. (c) Floating fitting range analysis in pure and Yb-substituted CeCoIn$_5$ samples. The data were fit using a power-law function over the temperature range from base temperature to a temperature $T_{up}<T_c/3$, and the resultant exponent $n$ was plotted as a function of $T_{up}$. }
\end{figure*}

In Fig.~\ref{fig3_doping} we summarize the evolution of the temperature-dependent London penetration depth in Ce$_{1-x}$R$_x$CoIn$_5$ ($R=$La, Nd, Yb). In panel (a) we show data for $R$=La, data for Nd;  Yb substitutions are shown in panel (b). The data are plotted versus a normalized temperature scale $(T/T_c)^2$. For reference we include pure CeCoIn$_5$, S1. As expected, the pure material shows downward curvature consistent with $n<$2. Doping with both La and Nd suppresses $T_c$ by as much as 0.5~K and 0.9~K respectively (see Fig.~\ref{fig1_samples}) and rapidly changes the power law exponent $n$ to 2 for $x$=0.05, as expected for line-node superconductors.
On the other hand, the evolution of the temperature-dependent London penetration depth with Yb doping is unique. The sample with $x$=0.2 demonstrates clear saturation at low temperatures, consistent with $n>$2, which cannot be explained in the nodal scenario. 
The increase of the exponent $n > 2$ can also be clearly seen in samples with $x$=0.1. 

As CeCoIn$_5$ is a multi-gap system\cite{Grenoble}, we must be careful in our analysis. In single gap $s$-wave and $d$-wave superconductors, the characteristic behavior of $\Delta \lambda (T)$ is observed for temperatures $T<T_c/3$, where the temperature dependence of the gap $\Delta (T)$ can be neglected. This assumption is not valid for multi-band systems, in which the smallest of the gaps can be strongly temperature-dependent down to temperatures $T<0.3T_c$. Since the range over which the smaller gap can be considered as constant is not known {\it a priori}, it is important to vary the range of the power-law fitting. We adopted a procedure in which the high temperature end of the fitting interval, $T_{up}$, was varied and the exponent $n$ of the power-law fit was plotted as a function of $T_{up}$, as shown in panel (c) of Fig.~\ref{fig3_doping}. Several conclusions can be drawn from the inspection of $n (T_{up})$ and its evolution with Yb doping. The dependence in samples 
 with $x$=0.1 and $x$=0.2 shows that the data are inconsistent with the existence of nodes in the superconducting energy gap for any temperature range selection, as for all $T_{up}$, $n > 2$, the highest value possible in superconductors with line nodes. Moreover, the exponent in the highest doped sample attains values which are technically indistinguishable from the exponential behavior observed in full gap superconductors \cite{Prozorov2006}. Hence, we conclude that the superconducting energy gap in CeCoIn$_5$ undergoes a topological transition from nodal to nodeless with Yb-substitution.
%Importantly, despite significant change of the gap structure, the superconducting $T_c$ decreases smoothly with $x$, see Fig.~\ref{fig1_samples}. 

%%%%%%%%%%%%% figure 4: doping evolution summary
\begin{figure}
\centering
\includegraphics[totalheight=2.7in]{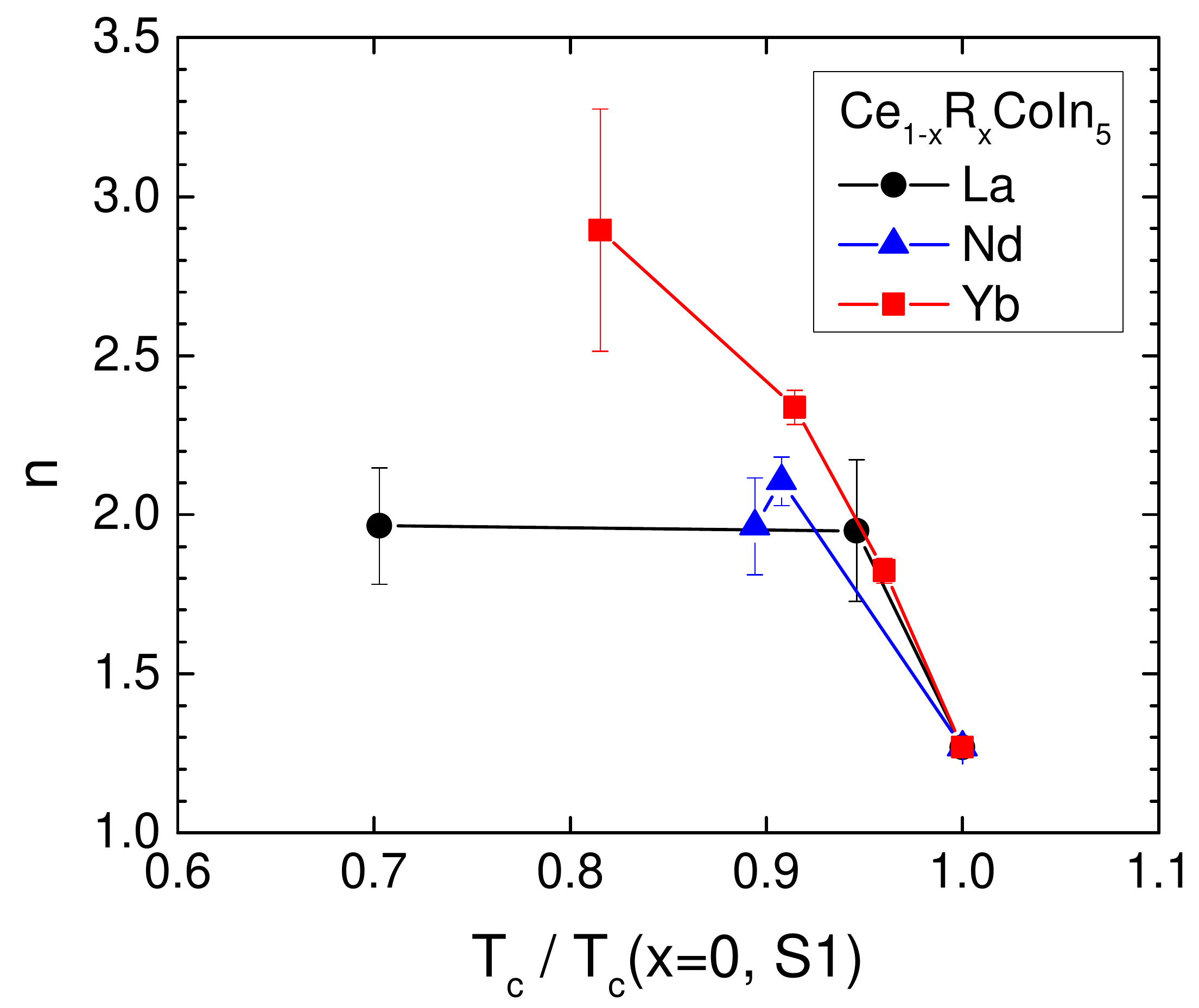}
\caption{\label{fig4_ntc} 
The average exponent of the power-law dependence of London penetration depth as a function of superconducting transition temperature for La substitution (black circles), Nd substitution (blue triangles) and Yb substitution (red squares). The values are obtained by the averaging of the exponent obtained from the range fitting, see Fig.~\ref{fig3_doping}(c).
}
\end{figure}

We summarize our study of doping evolution of London penetration depth and $T_c$ in rare earth substituted CeCoIn$_5$ in Fig.~\ref{fig4_ntc}.  We plot the exponent of the power-law analysis of London penetration depth as a function of $T_c$, with dopant concentration as a hidden parameter. This type of plot avoids uncertainty in the concentration of La, Nd and Yb substitutions. In particular, the rate of depression of $T_c$ with $x$ for Yb substituted flux grown singe crystals \cite{MapleYb} such as those studied here is only about one third of that found in thin film specimens \cite{MatsudaYb}. The origin of this difference is currently under investigation.

This study shows that Yb substitution significantly changes the structure of the superconducting energy gap in CeCoIn$_5$. Comparison to La and Nd substitutions suggests that this effect is not due to pairbreaking impurities or to doping-induced magnetism.  Instead, we conclude that the hole-doping effect of Yb substitution, and the resulting change in electronic structure is the important factor.  It is thus natural to link the change of the superconducting energy gap structure to a change in Fermi surface topology, and indeed recent de Haas-van Alphen studies find significant changes of the Fermi surface with Yb substitution, including the disappearance of the intermediately heavy $\alpha$ sheet between $x = 0.1$ and $0.2$\cite{Polyakov,Dudy}, exactly where we find that the gap nodes vanish.

STM studies of CeCoIn$_5$ \cite{Allan} indicate that the $\alpha$ Fermi surface sheet plays a key role in superconductivity. A change in the gap structure with its disappearance seems plausible, as the spectrum of magnetic fluctuations may change dramatically. However, two features in our data are difficult to reconcile with this scenario. First, $T_c$ evolves smoothly during the gap structure transformation, seemingly completely insensitive to the Fermi surface topology change. Second,  if the Fermi surface structure determines the gap structure, a symmetry-imposed d-wave gap is difficult to understand.

The iron-based superconductors have similar changes in superconducting energy gap structure: in hole-doped KFe$_2$As$_2$, a change in gap structure with pressure is indicated by a non-monotonic $T_c(P)$ dependence \cite{Fazel}, and similar changes with doping have been suggested in Ba$_{1-x}$K$_x$Fe$_2$As$_2$ \cite{ReidSUST}.  The gap structure of iron pnictides is frequently discussed in terms of Fermi surface nesting favoring magnetically mediated pairing, and similar ideas have been discussed for the 115s\cite{Ronning,TDas}.  However, in both iron-based examples, $T_c$ varies non-monotonically through the change, as one would expect for nesting dependent pairing, while in Yb doped CeCoIn$_5$, $T_c$ barely notices the change in gap structure, suggesting that the Fermi surface may not play an important role here.  One alternate scenario is that the superconducting energy gap symmetry remains d-wave through $x = 0.2$, but the underlying Fermi surface disappears - in this sense, the nodes in pure CeCoIn$_5$ are accidental.  Indeed, hole-doping should remove the $\alpha$ sheet seen in STM\cite{Allan}.  However, the removal of the Fermi surface sheet with the largest gap is problematic for magnetically mediated pairing theories.

%%%%%%%%%%%%%%%%%%%%%%%%%%%%%%%%%%%%%% Short CP paragraph

Composite pairing provides an alternate scenario\cite{Onur}.  Here, superconductivity arises from cooperative Kondo screening, where two electrons screen the same local moment to form a composite pair\cite{Piers}.  This process is local and does not require an underlying Fermi surface, allowing the Yb doping to tune the underlying heavy Fermi liquid toward a Kondo insulator without affecting the pairing.  The low energy excitations are uncondensed mobile composite pairs expected to have a $\Delta \lambda \sim T^4$ temperature dependence, consistent with our data for $x = 0.2$.

%%%%%%%%%%%%%%%%%%%%%%%%%%%% CONCLUSIONS

In conclusion, by performing systematic measurements of the London penetration depth in Ce$_{1-x}R_x$CoIn$_5$, $R$=La, Nd and Yb,  we find an anomalous change of the superconducting energy gap structure in Yb - doped compounds from nodal to nodeless linked with the Fermi surface topology change for $x = 0.2$.

%%%%%%%%%%%%%%%%%%%%%%%%%%%% ACKNOWLEDGMENTS

We thank P. Coleman and O. Erten for stimulating discussions and sharing their theoretical work prior to publication. The work in Ames was supported by the U.S. Department of Energy (DOE), Office of Science, Basic Energy Sciences, Materials Science and Engineering Division. Ames Laboratory is operated for the U.S. DOE by Iowa State University under contract No. DE-AC02-07CH11358. Part of the work was carried out at the Brookhaven National Laboratory, which is operated for the US Department of Energy by Brookhaven Science Associates (DE-Ac02-98CH10886).  Research at UCSD was supported by the U. S. DOE under Grant No. DE-FG02-04ER46105. H.~K. acknowledges the support from AFOSR-MURI grant No. FA9550-09-1-0603.

%%%%%%%%%%%%%%%%%%%%%%%%%%%% BIBLIOGRAPHY

\end{document}